\documentclass[amsmath,amssymb,aps,prl,reprint,superscriptaddress]{revtex4-1}
\usepackage{graphicx}
\usepackage{xcolor}
\usepackage[colorlinks=true,urlcolor=blue,citecolor=blue,linkcolor=blue]{hyperref}

\usepackage{physics}
\usepackage{siunitx}
\usepackage{xspace}
\usepackage{bm}
\usepackage{comment}

\newcommand{\vect}[1]{\vb{#1}}

\newcommand{\bk}{\textbf{k}}
\newcommand{\bq}{\vect{q}}

\usepackage[normalem]{ulem}

\definecolor{MAB-color}{named}{green}
\definecolor{AJ-color}{named}{blue}
\definecolor{ACG-color}{rgb}{0.97,0.57,0.11}
\definecolor{GB-color}{RGB}{128,0,128}

\definecolor{MAB-color2}{named}{green}
\definecolor{AJ-color2}{named}{red}
\definecolor{ACG-color2}{rgb}{0.87,0.47,0.01}
\definecolor{GB-color2}{RGB}{128,0,128}

\begin{document}


\title{Nonlinear optical response of resonantly driven polaron-polaritons}



\newcommand{\affiliationAarhus}{Center for Complex Quantum Systems, Department of Physics and Astronomy, Aarhus University, Ny Munkegade, DK-8000 Aarhus C, Denmark}
\newcommand{\affiliationMexico}{Departamento de F\'isica, Universidad Aut\'onoma Metropolitana-Iztapalapa, San Rafael Atlixco 186, C.P. 09340 CDMX, Mexico}
\newcommand{\affiliationCambridge}{T.C.M. Group, Cavendish Laboratory, University of Cambridge, JJ Thomson Avenue, Cambridge, CB3 0HE, U.K}
\newcommand{\affiliationChina}{Shenzhen Institute for Quantum Science and Engineering and Department of Physics, Southern University of Science and Technology, Shenzhen 518055, China}

\author{Aleksi Julku}
\affiliation{\affiliationAarhus}
\author{Miguel. A. Bastarrachea-Magnani}
\affiliation{\affiliationAarhus}
\affiliation{\affiliationMexico}
\author{Arturo Camacho-Guardian}
\affiliation{\affiliationCambridge}
\author{Georg Bruun}
\affiliation{\affiliationAarhus}
\affiliation{\affiliationChina}


\date{\today}

\begin{abstract}
Exciton polaritons in two-dimensional semiconductors inside microcavities are powerful platforms to explore hybrid light-matter quantum systems. 
Here, we study a macroscopic coherent population of the lowest energy state of polaron-polaritons, which are quasiparticles formed by the dressing of exciton polaritons by particle-hole excitations in a surrounding electron gas. 
 Using a non-perturbative many-body theory to describe   exciton-electron 
correlations combined with a non-equilibrium theory for the macroscopically  populated state,
we show that the electrons strongly affect the collective properties of the polaron-polaritons. This stems from the dependence of the polaron-polariton energy and the interaction between them on the electron density, which leads 
to strong nonlinearities. We identify stable and unstable regimes of the polaron-polaritons
by calculating its excitation spectrum,  and show that they result in prominent hysteresis effects when the electron density is varied. Our results should be readily observable using present  experimental technology. 
\end{abstract}


\maketitle

Exciton-polaritons featuring a coherent superposition of light and matter have gained widespread attention for exhibiting several interesting effects such as Bose-Einstein condensation and superfluidity~\cite{Kasprzak2006,Deng2010,Amo2009,Kohnle2011,Lagoudakis2008}, and topological photonics~\cite{Ozawa2019,Karzig2015,St-Jean2017,Klembt2018}. Recently, systems comprising exciton-polaritons interacting with an electron gas in microcavity semiconductors, such as monolayer transition metal dichalcogenides (TMDs), have been explored both theoretically and experimentally~\cite{Sidler2016,Efimkin2020,Bastarrachea-Magnani2020,Strashko2020,Shahnazaryan2020,Tan2020,Emmanuele2020}. An exciting perspective of these hybrid light-matter systems is the realisation of  strong optical nonlinearities~\cite{Tan2020,Emmanuele2020} with  applications in optoelectronics~\cite{Schaibley2016,Wang2018}.

 When an electron gas is injected into a semiconductor, it can  interact with the excitons to form exciton polarons~\cite{Efimkin2018,Chang2018}. Coupling to  cavity photons in turn makes them into polaron-polaritons~\cite{Sidler2016,Tan2020,Emmanuele2020,ExcitonNote}. While focus so far has mostly been on the limit of low polariton density,
these systems can realise other interesting regimes of physics as well. Depending on the ratio 
between the polariton and electron densities, this includes a gas of trions, i.e. bound-states of excitons and electrons~\cite{Rapaport2001,Efimkin2017,Kyriienko2020,Glazov2020,Rana2020}, electrons dressed by polaritons forming electron polaron-polaritons, and  Bose-Fermi mixtures~\cite{Cotlet2016}. Indeed, 
experiments realising atomic Bose-Fermi  mixtures have shown that they 
exhibit rich physics  such as
mediated interactions~\cite{DeSalvo2019}, phase separation~\cite{Lous2018}, mixed superfluids~\cite{FerrierBarbut2014}, and the breakdown of the Fermi polaron~\cite{Fritsche2021}. It is therefore interesting to explore such mixtures in the  hybrid light-matter setting of  polaritons, which  offer new possibilities for probing and control.

\begin{figure*}
  \centering
    \includegraphics[width=1.0\textwidth]{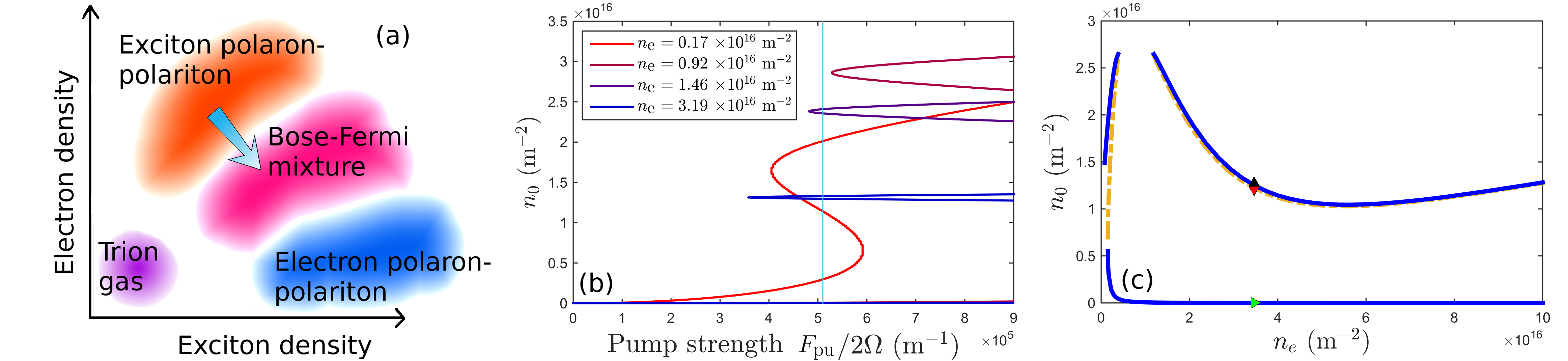}
    \caption{(a) Schematic diagram illustrating the different physical regimes realised by polaron-polaritons in a 2DEG. The arrow represents our approach to analyse the Bose-Fermi mixture, using the  polaron-polaritons as our starting point. (b) Condensate density $n_0$ as a function of pump laser amplitude $F_{\textrm{pu}}$ for the pump frequency $\epsilon_{\textrm{pu}}/2\Omega=-2.22$ and various electron densities $n_e$. (c)  Condensate density as a function of electron density $n_e$ for fixed pump laser amplitude    indicated by the vertical line in (b).  Blue solid (yellow dashed) lines indicate stable (unstable) configurations.}
   \label{Fig:1}
\end{figure*}

Motivated by this, we investigate here the case where 
a pump laser creates a macroscopic coherent population of polaritons immersed in an electron gas. Combining  a many-body theory of single polaron-polaritons 
with Gross-Pitaevskii and Bogoliubov theories, 
 we  show that not only the single particle but also the \emph{collective} properties of the polaritons are dramatically affected by the electrons.  This is because both the energy and the effective interaction between the polaron-polaritons depend on the electron density.
 Our theoretical approach is illustrated in  Fig.~\ref{Fig:1}(a).
We demonstrate that  one can drive the condensate into stable and unstable regimes with very different densities resulting in prominent bi-stabilities as shown in Fig.~\ref{Fig:1}(b)-(c), which have no analogue for bare polaritons. 

\textit{System}.---  
We study polaritons created in a two-dimensional semiconductor in a pump-probe experiment, where  excitons strongly couple to a cavity light field. In addition, they interact with each other as well as with an electron gas (2DEG). The excitons are treated as a point bosons due to their large binding energy in  semiconductors such as the TMDs~\cite{Bleu2020,Wang2018}. The Hamiltonian is 
\begin{gather}
H = \sum_{\bk} \begin{bmatrix} \hat{x}_{\bk}^\dag & \hat{c}_{\bk}^\dag \end{bmatrix}
\begin{bmatrix}
\epsilon_{\bk}^{(x)} & \Omega \\ \Omega & \epsilon_{\bk}^{(c)} 
\end{bmatrix}
\begin{bmatrix}
\hat{x}_{\bk} \\ \hat{c}_{\bk}
\end{bmatrix} + \sum_{\bk} \epsilon_{\bk}^{(e)} \hat{e}^\dag_{\bk} \hat{e}_{\bk}+ \nonumber \\
  \sum_{\bk,\bk',\bq}\left(g_{ex}\hat{x}_{\bk'-\bq}^\dag \hat{e}_{\bk+\bq}^\dag \hat{e}_{\bk} \hat{x}_{\bk'}+
\frac{g_{xx}}2\hat{x}_{\bk'-\bq}^\dag \hat{x}_{\bk+\bq}^\dag \hat{x}_{\bk} \hat{x}_{\bk'}\right),
\label{ham1}
\end{gather}
where $\hat{x}_{\bk}$, $\hat{c}_{\bk}$ and $\hat{e}_{\bk}$ are the annihilation operators for an exciton, cavity 
photon and electron, respectively, with energies $\epsilon_{\bk}^{(x)} = \bk^2/2m_x$, $\epsilon_{\bk}^{(c)} = \bk^2/2m_c +\delta$ and $\epsilon_{\bk}^{(e)} = \bk^2/2m_e -\mu_e$, and masses $m_x$, $m_c$, and $m_e$. The  exciton-photon coupling strength is $\Omega$, and  the electron-exciton $g_{ex}$ and exciton-exciton $g_{xx}$ interactions are taken to be momentum independent due to their short range nature. We use units where the system area and $\hbar$ are one, 
and the energy offset of the electrons relative to the excitons is 
absorbed in their chemical potential $\mu_e$.
Due to the optical and valley selection rules of TMDs~\cite{Wang2018}, photons couple to excitons in a given valley with a fixed spin state, which in turn interact predominantly with electrons in another valley. We can therefore ignore the valley and spin index of the excitons, electrons, and photons.

\begin{figure}
  \centering
    \includegraphics[width=1.0\columnwidth]{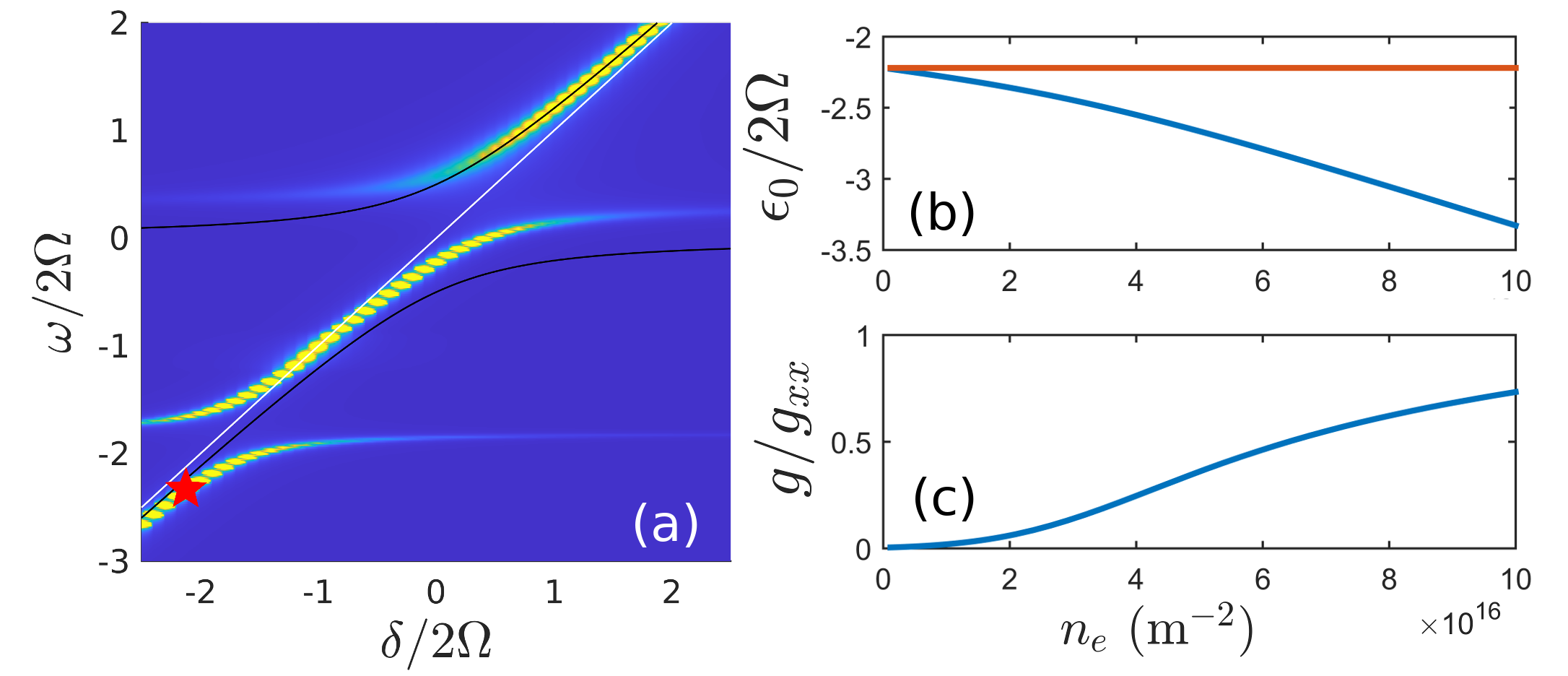}
    \caption{(a) Spectral density of a single cavity photon. White line is the bare photon energy and black lines are the bare upper and lower polariton dispersions. The energy $\epsilon_0$ of the lowest polaron-polariton branch at $\delta/2\Omega =-2.11$ is indicated by a red star 
    (b) The energy $\epsilon_0$  as a function of the electron density. The pump frequency at $\epsilon_{\textrm{pu}}/2\Omega=-2.22$ is shown as a horizontal line.
    (c) The  interaction between two polaron-polaritons as a function of the electron density.}
   \label{Fig:2}
\end{figure}

In the absence of interactions, diagonalization of Eq.~\eqref{ham1} yields the  energies $\epsilon^{LP/UP}_\bk = \frac{1}{2}( \epsilon_{\bk}^{(c)} + \epsilon_{\bk}^{(x)} \pm \sqrt{\left( \epsilon_{\bk}^{(c)} - \epsilon_{\bk}^{(x)}\right)^2 + 4\Omega^2})$ of the lower (LP) and upper (UP) polariton~\cite{Hopfield1958}.
They are depicted in Fig.~\ref{Fig:2}(a) as black lines for $\bk=0$ as a function of the light-matter detuning $\delta$.

\textit{Polaron-polaritons}.---
The interaction between the excitons and electrons in the presence of strong light coupling has been observed to lead to the formation of polaron-polaritons~\cite{Sidler2016,Tan2020,Emmanuele2020}, in analogy with 
Fermi polarons  in atomic gases~\cite{Massignan2014}. A convenient way to describe them is to introduce a $2\times 2$ matrix Green's function, which in Matsubara frequency space reads~\cite{Bastarrachea-Magnani2020,SM}
\begin{align}
\label{g_ind}
\mathcal G^{-1}(\bk,i\omega_n) = \begin{bmatrix} i\omega_n - \epsilon^{(x)}_{\bk} &0 \\ 0 & i\omega_n - \epsilon^{(c)}_{\bk} \end{bmatrix}-\begin{bmatrix}  \Sigma_{x}(\bk,i\omega_n) & \Omega \\ \Omega & 0 \end{bmatrix}
\end{align}
where the effects of the strong exciton-electron interaction  are included via the exciton self-energy 
\begin{align}
\label{selfenergy}
\Sigma_{x}(\bk,i\omega_n)=\int \frac{d^2\bq}{(2\pi)^2}n_F(\epsilon^{(e)}_{\bq})\mathcal{T}(\bk+\bq,i\omega_n + \epsilon^{(e)}_{\bq}).
\end{align}
Here, $\omega_n$ is a bosonic Matsubara frequency and $n_F(x)=(\exp\beta x+1)^{-1}$ is the Fermi-Dirac distribution function. The exciton-electron scattering matrix is ${\mathcal T}(\bk,i\omega_n)$, which in  the ladder approximation reads
$\mathcal{T}^{-1}(\bk,i\omega_n)=\mbox{Re}\Pi_{V}(\epsilon_B)-\Pi(\bk,i\omega_n)$ with
$\Pi$ and $\Pi_{V}$ being the exciton-electron pair propagator in the presence of the 2DEG and in a vacuum, respectively~\cite{SM,Bastarrachea-Magnani2019,Wouters2007,Carusotto2010}. The exciton-electron interaction is strong enough to support a bound state, a trion, giving rise to a pole of the scattering matrix at its energy, which is shifted away from its vacuum value $\epsilon_B$ due to medium effects.

In Fig.~\ref{Fig:2}(a) we show the photon spectral function  $A_c(\bk,\omega)$, obtained by diagonalizing Eq.~\eqref{g_ind}, for zero momentum $\bk=0$ and electron density of $n_e = 1.58 \times 10^{16}$ m$^{-2}$. Here, and in the rest of the paper, we take $m_x=2m_e$, $m_c = 10^{-5}m_e$, and use an experimentally inspired value for the trion energy $\epsilon_B/2\Omega= -1.56$ \cite{Tan2020,Sidler2016,Mak2012} with $\Omega=8$ meV. Furthermore, we focus on the zero-temperature limit so that $\mu_e = \epsilon_F$, with $\epsilon_F$ being the Fermi energy. Figure \ref{Fig:2}(a) exhibits three quasiparticle branches, obtained by solving $\text{Re}\left[\det\mathcal G^{-1}(\bk,\omega_{\bk} )\right]=0$, with energies  
\begin{align}
\label{pol-pol-en}
\omega_{\bk} =& \frac{1}{2}\Big[ \epsilon^{(c)}_{\bk} + \epsilon^{(x)}_{\bk} + \Sigma_x(\bk,\omega_{\bk}) 
\pm \sqrt{{\delta}_{\bk}^{2} + 4\Omega^2} \Big].
\end{align}
Here, the detuning ${\delta}_\bk = \epsilon^{(c)}_{\bk} - \epsilon^{(x)}_{\bk}-\Sigma_x(\bk,\omega_{\bk})$ includes many-body effects and carries momentum dependence. Depending on the values of $\delta$ and $\epsilon_F$, Eq.~\eqref{pol-pol-en} with either $+\sqrt{\ldots}$ or $-\sqrt{\ldots}$
 has two solutions. This gives rise to three polaron-polariton  branches in Fig.~\ref{Fig:2}(a), which differ significantly from the bare polariton branches. 

From now on, we focus on the lowest energy $\bk=0$ polaron-polariton  for $\delta/2\Omega=-2.11$ 
indicated by the red star in Fig.~\ref{Fig:2}(a). In Fig.~\ref{Fig:2}(b), we plot its energy $\omega_{\bk=0} \equiv \epsilon_0$ as a function of the electron density $n_e$. Importantly, we see that it decreases with increasing $n_e$ as it gets increasingly repelled by the trion state. 

\textit{Polaron-polariton interaction.---}
 The effective interaction $g$ between zero-momentum polaron-polaritons, responsible for the non-linear optical properties, is modified by the coupling to the 2DEG. It is given by
\begin{align}
g = g_{xx} {C}^4(\bk=0) Z^2(\bk=0)
\label{EffectiveInt}
\end{align}
where ${C}^2(\bk) = 1/2\big(1 +{\delta}_\bk/\sqrt{{\delta}^2_\bk + 4\Omega^2}\big)$ is the generalised Hopfield coefficent, which takes into account that 
 polaron-polaritons interact only via their excitonic component. Compared to the usual Hopfield coefficients~\cite{Hopfield1958}, it includes many-body effects via the detuning ${\delta}_{\bk}$. 
The  residue $Z^{-1}(\bk)=1-{C}^2(\bk)\partial_\omega\left.\Sigma_{x}(\bk,\omega)\right|_{\epsilon_0}$ describes the fact that many-body correlations reduce the plane wave component of the polaron-polariton as compared to the bare polaritons and is 
typical for microscopic expressions of effective interactions between quasiparticles~\cite{Giuliani2005,Camacho2018}.

To analyse the effects of the 2DEG on the interaction between the polaron-polaritons, we plot in Fig.~\ref{Fig:2}(c) $g$ as a function of $n_e$. It shows that the interaction 
 increases significantly with the electron density. The reason is that the electrons decrease the polaron-polariton energy, Fig.~\ref{Fig:2}(b), which increases ${\delta}_\bk$ and therefore the excitonic component of the polaron-polaritons given by ${C}(\bk)$. The two effects shown in Fig.~\ref{Fig:2}, the decrease in the polaron-polariton energy and the increase in their  interaction,  play a key role for the non-linear effects we present later. 

Determining the value of the bare exciton-exciton interaction strength $g_{xx}$ remains a challenging problem, and there are several discrepancies between theory and experiment. While it has been estimated that $g_{xx}\sim{\mathcal O}(10)\mu$eV$\mu$m$^2$
in GaAs quantum wells~\cite{Ciuti1998,Ferrier2011,Rodriguez2016,Delteil2019}, theoretical calculations predict $g_{xx}\sim 2\mu$eV$\mu$m$^2$ in  TMDs~\cite{Shahnazaryan2017}, whereas experimental estimates range from $g_{xx}\sim 0.05\mu$eV$\mu$m$^2$~\cite{Barachati2018,Tan2020}  to $g_{xx}\sim 3.0\mu$eV$\mu$m$^2$~\cite{Emmanuele2020}. In this work we choose $g_{xx}= 3\mu$eV$\mu$m$^2$. Using a smaller interaction would yield larger condensation densities without changing the results qualitatively.

There is also an attractive interaction between the  excitons mediated by the 2DEG~\cite{Bastarrachea-Magnani2020} as well as a repulsive non-equilibrium phase filling effect~\cite{Tan2020}. Both can safely be neglected here since we consider steady-state properties focusing on the regime where the energy $\epsilon_0$ of the lowest polaron-polariton  is well below that of the trion. 

In principle, the presence of polarons can lead to mediated superconductivity of the 2DEG. However, this effect should be small as the density of polarons is assumed to be smaller than the density of 2DEG. Furthermore, the spin-polarised character of the electrons prevents $s$-wave pairing, hence, possible exotic electronic phases such as $p$-wave superconductivity are even further suppressed. Since the p-wave gap is predicted to be much smaller than the Fermi energy, which in turns is smaller than the light-matter coupling, superconductivity is expected to have negligible effects on the properties of the polaron-polaritons, which is the focus of the present paper.

\textit{Polaron-polariton condensate}.--- 
We are now ready to analyse the collective properties of polaron-polaritons. The system is assumed to be driven coherently by a continuous pump laser  with zero angle incidence, which introduces a macroscopic occupation to the $\bk=0$ ground state of the polaron-polaritons with energy 
$\epsilon_0$, marked by the red star in Fig.~\ref{Fig:2}(a). In the following, this is for brevity referred to as a condensate. 

We will use the single particle properties calculated above as a starting point, even though the theory used is strictly valid only for a single polaron-polariton. For atomic gases, such an approach has turned out to be accurate for surprisingly high polaron concentrations~\cite{Lobo2006,Fritsche2021} and we expect the same to be the case for the polaron-polaritons studied here. 
The driven-dissipative nature of semi-conductor microcavity systems  is known to give rise to qualitatively new effects for polariton condensates.
 We will therefore employ a non-equilibrium 
Gross-Pitaevskii equation (GPE)
to describe the  behaviour of the  polaron-polariton condensate. It reads~\cite{Carusotto2004,Ciuti2005,Carusotto2013,SM}
\begin{align}
\label{gpe}
[\epsilon_{\textrm{pu}} - \epsilon_0(n_e) - g(n_e) |\Psi|^2  + i\gamma]\Psi = iF_{\textrm{pu}},
\end{align}
The condensate wave function is denoted by $\Psi,$ while   $\epsilon_{\textrm{pu}}$ is the pump laser frequency, and $F_{\textrm{pu}}$ is its amplitude.
We use an experimentally realistic value $\gamma=0.4$ meV for the 
polaron-polariton loss rate due to processes such as exciton decay and photon leakage through the cavity mirrors~\cite{Emmanuele2020}. Note that in Eq.~\eqref{gpe} we have explicitly expressed the dependence of $\epsilon_0$ and $g$ on $n_e$ to highlight that these are the parameters being varied when the electron density is tuned as shown in Fig.~\ref{Fig:2}. 
The pump laser frequency is $\epsilon_{\textrm{pu}}/2\Omega=-2.22$ as indicated by the horizontal line in Fig.~\ref{Fig:2}(b), which shows that the laser becomes increasingly blue detuned with respect to the polaron-polaritons as $n_e$ gets larger.

 In Fig.~\ref{Fig:1}(b), we plot the condensate density $n_0=|\Psi|^2$ obtained from Eq.~\eqref{gpe} as a function of the pump laser amplitude $F_{\textrm{pu}}$ for different electron densities. We see that as the electron density increases, an s-shape feature emerges, which is characteristic of hysteresis effects. The height of the s-shape  first increases and then decreases with $n_e$. This non-trivial behavior  is due to the combination of an increasing blue detuning of the laser and the growth of the interaction strength with the increase of the electron density.
 
 To further investigate this, we plot in Fig.~\ref{Fig:1}(c) the condensate density as a function of the electron density for a fixed pump laser amplitude  indicated by the vertical line in Fig.~\ref{Fig:1}(b). This shows that the s-shaped curves in Fig.~\ref{Fig:1}(b) yield three solutions for low and high electron densities $n_e$. The behavior for low $n_e$ arises predominantly due to the increasing blue detuning of the laser with increasing $n_e$. This is similar to what can be observed when the laser frequency $\epsilon_{\textrm{pu}}$ is tuned for polariton condensates without the 2DEG~\cite{Carusotto2004,Ciuti2005,Carusotto2013}. 
 
 The behavior for large electron density on the other hand, arises due to an intriguing interplay between an increasing blue detuning of the pump  and an increasing coupling strength \cite{SM}. Physically, the presence of the 2DEG leads to a red-shift of the polaron-polariton energy, which is accompanied by an increase in its matter component. This in turn increases the coupling strength and these two effects combined give rise to prominent optical non-linearities. They have no analogue when the 2DEG is absent and are thus a  unique property of the Bose-Fermi mixture. We furthermore note that  this new bi-stability occurs for high electron densities where our theoretical approach is most reliable.

\begin{figure}
  \centering
    \includegraphics[width=1.0\columnwidth]{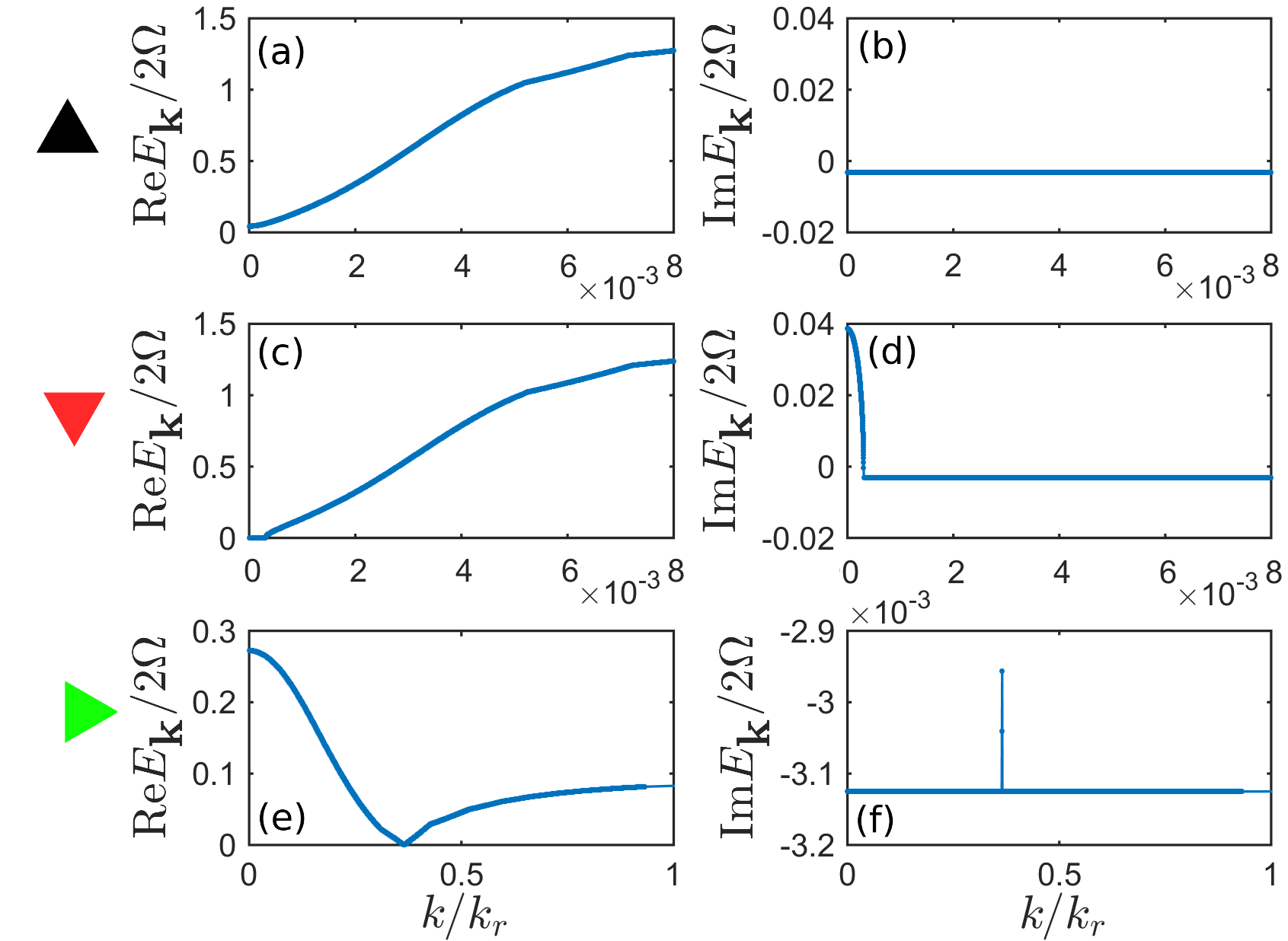}
    \caption{Bogoliubov dispersions $E_\bk$ as a function of the in-plane momentum $k =|\bk|$ for the three different cases marked by the triangles in Fig.~\ref{Fig:1}(c).  Panels (a), (c) and (e) show Re$[E_\bk]$ and panels (b), (d) and (f) show Im$[E_\bk]$. Momenta are expressed in the units of $k_r = \sqrt{4\pi n_0}$. The stability of different cases can be deduced from the sign of the imaginary part of $E_\bk$.}
   \label{Fig:3}
\end{figure}

\textit{Excitation spectrum}.--- 
To explore the stability of the solutions of the GPE, we calculate the excitation spectrum of the condensate of polaron-polaritons. As shown in the Supplemental material \cite{SM}, Bogoliubov 
theory  generalised to the non-equilibrium pump-loss setup yields 
 the 2$\times$2 boson Green's function 
\begin{align}
&\mathcal G^{-1}_\text{B}(\bk,\omega) = \begin{bmatrix} \omega - {\xi}_\bk-2gn_0 +i\gamma&-gn_0 \\ -gn_0 & -\omega -  {\xi}_\bk-2gn_0 -i\gamma  \end{bmatrix}
\end{align}
where $\xi_\bk=\varepsilon_\bk-\epsilon_{\textrm{pu}}$, with ${\varepsilon}_\bk$ being the energy of the lowest polaron-polariton obtained by solving Eq.~\eqref{pol-pol-en}.
 Note that  the single-particle energy in the hole-channel is conjugated since it contains the decay $\gamma$ \cite{SM}. The excitation energies $E_\bk$ can then be found as the poles of $\mathcal G_\text{B}(\bk,\omega)$, giving
\begin{align}
\label{bogo}
E_\bk = \sqrt{({\xi}_\bk + 2g n_0)^2 - (g n_0)^2} - i\gamma.  
\end{align}

Fig.~\ref{Fig:3} shows the excitation spectrum $E_\bk$ for the three cases marked by the coloured triangles in Fig.~\ref{Fig:1}(c). The top panels show the case marked by the black triangle 
in Fig.~\ref{Fig:1}(c) corresponding to $n_e = 3.5\times 10^{16}$ m$^{-2}$ and $n_0= 1.25\times 10^{16}$ m$^{-2}$. We see that the  spectrum is gapped, which is  because the pump laser breaks the  $\text{U}(1)$ phase symmetry of the wave funcion $\Psi$  in contrast to the  gapless Goldstone mode  of equilibrium condensates~\cite{Ciuti2005,Carusotto2013}. 
Moreover, the energies have a negative imaginary part   for all $\bk$, meaning that the condensate is stable. 

Figures \ref{Fig:3}(c)-(d) depict the case marked by the red  triangle 
in Fig.~\ref{Fig:1}(c) with $n_e = 3.5\times 10^{16}$ m$^{-2}$ and $n_0= 1.22\times 10^{16}$ m$^{-2}$. The real part of $E_\bk$ is exactly zero and the imaginary part is positive for small momenta. This feature emerges when the square root in Eq.~\eqref{bogo} becomes negative for  $-3gn_0 < {\textit{}}_\bk < -gn_0$, giving a purely imaginary contribution to the energy.  For small $\mathbf{k}$, this is larger than  the loss rate $\gamma$, making the system  dynamically unstable. 

Finally, the spectrum shown in Figs.~\ref{Fig:3}(e)-(f) is for the case marked by the green triangle 
in Fig.~\ref{Fig:1}(c) with $n_e = 3.5\times 10^{16}$ m$^{-2}$ and $n_0= 3.51\times 10^{12}$ m$^{-2}$. Here, the spectrum is qualitatively  different with a low momentum regime exhibiting a gapless mode at a finite momentum, implying a possible instability towards a supersolid phase. The effect of the 2DEG is, however, not large enough to render the system unstable. 

By analysing the stability of all the solutions of the GPE shown in Fig.~\ref{Fig:1}(c), we can identify the stable and unstable regions of the condensate, which are marked by blue/red lines respectively. These stable/unstable regions 
 give rise to an interesting hysteresis behaviour of the condensate and demonstrate in a dramatic way how the 2DEG can affect the collective properties of the polaritons.


\textit{Outlook and conclusions}.--- 
We explored the collective properties of polaron-polaritons in the presence of a 2DEG. Using a non-perturbative many-body theory, we first showed that the electrons affect both the energy of the polaron-polaritons and the interaction between them. The interplay of these two effects were then demonstrated to lead to a highly non-linear behaviour of the condensate, which was analysed using a non-equilibrium Bogoliubov theory. In particular,  the condensate exhibits bi-stabilities resulting in prominent hysteresis effects
that have no analogue in the absence of the 2DEG. This 
shows that the electron gas not only affects the single particle properties of the polaritons but also their collective behavior. 

Polaritons have already been observed to exhibit non-linear behavior when the pump laser frequency is varied~\cite{Baas2004,Ballarini2013}, and one has furthermore realised polaron-polaritons in the presence of a 2DEG~\cite{Rapaport2001,Bajoni2006,Sidler2016}.  
Thus, the  effects presented here  should be within reach of present day experimental technology by combining such experiments. From a broader perspective, our results demonstrate that  polaritons immersed in an electron gas can realise novel hybrid light-matter Bose-Fermi mixtures with interesting properties. This motivates future 
investigations into such mixtures where optical non-linearities have been observed~\cite{Tan2020,Emmanuele2020}
and strong mediated interactions between the polaron-polaritons predicted~\cite{Bastarrachea-Magnani2020,Camacho-Guardian2020}. Inspired by atomic gases, one may  observe phase separation~\cite{Lous2018}, strong coupling effects~\cite{Fratini2013,Guidini2015}, 
mediated topological and exotic superfluidity~\cite{Wu2016,Ozawa2014}. It is also an  interesting challenge to go beyond our polaron approach and develop a theory that accounts for strong coupling effects both on the polaritons and on the electrons, while at the same time describing the non-equilibrium features in a systematic way.

\begin{acknowledgments}
G.M.B.\ acknowledges financial support from DNRF through the Center for Complex Quantum Systems (Grant agreement No. DNRF156) and the Independent Research Fund Denmark-Natural Sciences (Grant No.\ DFF-8021-00233B). A.J. \ acknowledges financial support from the Jenny and Antti Wihuri Foundation. 
\end{acknowledgments}

\vspace{2cm}

\end{document}


\title{Nonlinear
optical response of resonantly driven polaron-polaritons: supplementary material}

\newcommand{\affiliationAarhus}{Department of Physics and Astronomy, Aarhus University, Ny Munkegade, DK-8000 Aarhus C, Denmark}
\newcommand{\affiliationMexico}{Departamento de F\'isica, Universidad Aut\'onoma Metropolitana-Iztapalapa, San Rafael Atlixco 186, C. P. 09340, Ciudad de M\'exico, M\'exico}
\newcommand{\affiliationCambridge}{T.C.M. Group, Cavendish Laboratory, University of Cambridge, JJ Thomson Avenue, Cambridge, CB3 0HE, U.K}
\newcommand{\affiliationChina}{Shenzhen Institute for Quantum Science and Engineering and Department of Physics, Southern University of Science and Technology, Shenzhen 518055, China}

\author{Aleksi Julku}
\affiliation{\affiliationAarhus}
\author{Miguel. A. Bastarrachea-Magnani}
\affiliation{\affiliationAarhus}
\affiliation{\affiliationMexico}
\author{Arturo Camacho-Guardian}
\affiliation{\affiliationCambridge}
\author{Georg Bruun}
\affiliation{\affiliationAarhus}
\affiliation{\affiliationChina}


\date{\today}


\maketitle

\onecolumngrid

\section{Exciton-exciton interactions}

\begin{figure}
  \centering
    \includegraphics[width=0.8\textwidth]{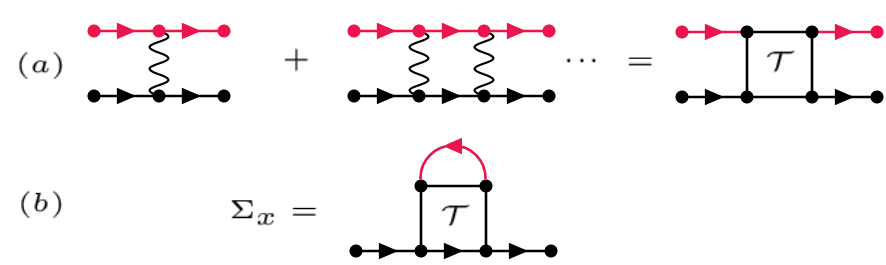}
    \caption{(a) Ladder approximation and the $\mathcal{T}$-matrix. Solid black (red) lines correspond to exciton (electron) propagators and wavy lines represent bare exciton-electron interaction vertices $g_{ex}$ (b) Exciton self-energy $\Sigma_x$.}
   \label{fig:ladder}
\end{figure}

The interaction Hamiltonian between electrons and excitons is 
\begin{align}
\label{interaction_h}
H_{int} = \sum_{\bk,\bk',\bq}g_{ex}\hat{x}_{\bk'-\bq}^\dag \hat{e}_{\bk+\bq}^\dag \hat{e}_{\bk} \hat{x}_{\bk'}.   
\end{align}
We base our approach on the Bethe-Salpeter equation considering the repeated forward scattering between an exciton and an electron as illustrated in Fig.~\ref{fig:ladder}(a). This so-called ladder approximation yields the $\mathcal{T}$-matrix~\cite{Fetter1971,Bastarrachea-Magnani2019}:
\begin{align}
\label{tmatrix1}
\mathcal{T}(\bk,i\omega_n) = g_{ex} + g_{ex} \Pi(\bk,i\omega_n)\mathcal{T}(\bk,i\omega_n)  \Leftrightarrow \mathcal{T}^{-1}(\bk,i\omega_n) = g_{ex}^{-1} - \Pi(\bk,i\omega_n),  
\end{align}
where $\omega_n = \frac{(2n+1)\pi}{\beta}$ are fermionic Matsubara frequencies ($n$ is an integer, $\beta =\frac{1}{k_B T}$, $k_B$ is the Boltzmann constant and $T$ is the temperature) and the exciton-electron pair-propagator $\Pi(\bk,i\omega_n)$ reads
\begin{align}
\Pi(\bk,i\omega_n) = -\frac{1}{\beta} \int \frac{d^2\bq}{(2\pi)^2} \sum_{iq_n}\mathcal{G}_x^0(\bk+\bq,iq_n) \mathcal{G}^0_e(-\bq,i\omega_n-iq_n).
\end{align}
Here $q_n = \frac{2\pi n}{\beta}$ are bosonic Matsubara frequencies and the non-interacting Green's functions for the excitons and electrons are, $\mathcal{G}^0_x(\bk,z) = 1/(z-\epsilon_{\bk}^{(x)})$ and $\mathcal{G}^0_e(\bk,z) = 1/(z-\epsilon_{\bk}^{(e)})$, respectively. 

In the limit of a single exciton-impurity, the pair propagator is 
\begin{align}
\Pi(\bk,i\omega_n) = \int \frac{d^2\bq}{(2\pi)^2} \frac{1- n_F(\epsilon_{\bq}^{(e)} - \mu_e)}{i\omega_n - \epsilon_{\bq}^{(e)} - \epsilon_{\bk+\bq}^{(x)}}.    
\end{align}
We use this form in the calculations as we are interested in the energy spectrum of a single polaron-polariton. 

In the vacuum limit, i.e. when a single exciton scatters with a single electron, the Bethe-Salpeter is an exact solution of the two-body problem. In this case, the two-body $\mathcal{T}$-matrix can be written as $\mathcal{T}_{V}^{-1}(0,i\omega_n) = g_{ex}^{-1} - \Pi_{V}(0,i\omega_n)$. In a two-dimensional system the two-body scattering problem always features a bound state of energy $-\epsilon_B<0$ \cite{Randeria1990} which manifests as a pole of $\mathcal{T}_{V}$ such that $g_{ex}^{-1} = \Re[\Pi_{V}(0,-\epsilon_B)]$. In case of the exciton-polaron problem, this bound state corresponds to the trion~\cite{Sidler2016,Mak2012,Tan2020,Bastarrachea-Magnani2020}. The many-body $\mathcal{T}$-matrix can then be recast in the renormalized form as
\begin{align}
\mathcal{T}^{-1}(\bk,i\omega_n) = \Re[\Pi_{V}(0,-\epsilon_B)] - \Pi(\bk,i\omega_n).    
\end{align}
The vacuum pair-propagator $\Pi_V$ reads
\begin{align}
\Pi_V(0,-\epsilon_B) = - \frac{m_e}{2\pi(1+\alpha)}\log\Big[\frac{(1+\alpha)\Lambda^2}{2m_e \epsilon_B}\Big],    
\end{align}
where $\Lambda$ is the ultraviolet momentum cut-off and $\alpha = m_e/m_x$. With large enough $\Lambda$, the ultraviolet divergences of $\Pi_V$ and $\Pi$ cancel each other and $\mathcal{T}$ is a well-defined function independent of $\Lambda$.

We employ a diagrammatic approach based on the $\mathcal{T}$-matrix, that takes into account the underlying trion-state. The self-energy for the excitons $\Sigma_x(\bk,i\omega_n)$ within the $\mathcal{T}$-matrix approximation is depicted in Fig.\ref{fig:ladder}(b), and given explicitly by
\begin{align}
\Sigma_x(\bk,i\omega_n) = \frac{1}{\beta}\int \frac{d^2\bq}{(2\pi)^2}\sum_{iq_n} \mathcal{G}^0_e(\bq,iq_n)\mathcal{T}(\bk+\bq,i\omega_n +iq_n).
\end{align}
Here $i\omega_{n}$ are now bosonic Matsubara frequencies. As we are considering the single exciton-impurity limit, $\Sigma_x$ simply reads
\begin{align}
\Sigma_x(\bk,i\omega_n) = \int \frac{d^2\bq}{(2\pi)^2} n_F(\epsilon_{\bq}^{(e)})\mathcal{T}(\bk+\bq,i\omega_n + \epsilon_{\bq}^{(e)}).   
\end{align}
The full Green's function for excitons is then $\mathcal{G}_x^{-1}(\bk,i\omega_n) = z - \epsilon^{(x)}_{\bk} - \Sigma_x(\bk,i\omega_n)$. By computing the energy poles of $\mathcal{G}_x$, one finds that $\Sigma_x$ gives rise to the attractive and repulsive Fermi-polaron states in the absence of light~\cite{Bastarrachea-Magnani2020}, in a similar fashion as in Fermi gases~\cite{Massignan2014,Schmidt2012}. 
The cavity photon field then couples to the polaron states giving rise to  three exciton polaron-polariton branches  as it is shown in Fig. 2(a) of the main text.

The emergence of exciton polaron-polaritons can be studied with the $2\times2$ Green's function describing both excitonic and photonic degrees of freedom. More precisely, we define $\mathcal{G}(\bk,\tau) = -\left\langle T_\tau \left\{\hat{\Psi}_{\bk}(\tau)  \hat{\Psi}^\dag_\bk(0)\right\} \right\rangle$, 
where $\hat{\Psi}_{\bk}^\dag \equiv  \begin{bmatrix}\hat{x}_{\bk}^\dag & \hat{c}_{\bk}^\dag  \end{bmatrix}$ 
and $T_\tau$ is the imaginary time-ordering operator. In the frequency space, we have~\cite{Bastarrachea-Magnani2020}
\begin{align}
\label{g_ind}
\mathcal{G}^{-1}(\bk,i\omega_{n}) = \begin{bmatrix} i\omega_{n} - \epsilon^{(x)}_{\bk} - \Sigma_{x}(\bk,i\omega_n) & -\Omega \\ -\Omega & i\omega_{n} - \epsilon^{(c)}_{\bk} \end{bmatrix},
\end{align}
which is the form used in the main text. 

%
%

\section{Non-equilibrium Bose condensation and Bogoliubov theory}

In this section, we provide details on the derivation of the non-equilibrium Gross-Pitaevskii equation (GPE) and the Bogoliubov equations giving the excitation spectrum.
To treat the incoherent decay terms of polaritons correctly, we deploy a formalism where the bosonic polariton modes are coupled to an external thermal reservoir bath of harmonic oscillators~\cite{Meystre2007,Scully1997}. The role of the reservoir is to take into account the open quantum system nature of the polaron-polariton setup. 
By assuming that the correlation times of the bath are much smaller than any relevant time scale of the polariton system, i.e. by employing a Markovian approximation, one ends up with the equations of motion which describe the decay of the polaritons.

We emphasize that the right form for the non-equilibrium Green's function can not be obtained with a non-hermitian decay Hamiltonian of the form $H_{decay} = \sum_\bk i\gamma \bc \ba$, where $\ba$ is the annihilation operator for the  polaron-polariton state at $\bk$. This can be seen from the Heisenberg equation of motion which  yields $\frac{d \ba}{dt} = -i[\ba,H_{decay}] = -\gamma \ba$ but  $\frac{d \bc}{dt} = \gamma \bc$. The solutions for these differential equations give a damping solution for $\ba$ but an unphysical amplifying solution for $\bc$. This inevitably leads to a wrong non-equilibrium theory. Below, we derive the Bogoliubov theory accounting correctly for the losses of the polaron-polaritons, preventing unphysical amplifications of the bosonic fields.

We start by writing down the effective Hamiltonian for the lower polaron-polaritons in the presence an external coherent driving field and incoherent losses:
\begin{align}
&H_{L} = H_{s} + H_{r} + H_{sr}, \textrm{ where} \nonumber\\
& H_s = \sum_\bk \epsilon_\bk \bc\ba +  \frac{g}{2A}\sum_{\bk,\bk',\bq} \hat{b}^\dag_\bk \hat{b}^\dag_{\bk'}  \hat{b}_{\bk'-\bq} \hat{b}_{\bk+\bq} + iF_{\pu}e^{-i\epsilon_{\pu}t} b^\dag_{\bk_{\pu}}, \nonumber \\
& H_{r} = \sum_j \epsilon^B_j \hat{r}_j^\dag \hat{r}_j \nonumber \\
& H_{sr} = \sum_{\bk,j} \big[ \kappa_j^* \bc \hat{r}_j + \kappa_j \hat{r}^\dag_j \ba \big].
\end{align}
Here the polaron-polaritons are described by the system Hamiltonian $H_s$, the reservoir bath is $H_r$ and the coupling between the system and the bath is treated via $H_{sr}$. The polaron-polariton energies are denoted by $\epsilon_\bk$, $A$ is the system area, $g$ is the effective interaction between polaron-polaritons, $F_\pu$ is the strength of the laser pump, $\epsilon_\pu$ is the pump energy and $\bk_\pu$ is the in-plane momentum of the pump ($\hbar =1$). In $H_r$, $\hat{r}_j$ are the annihilation operators for the $j$th reservoir mode of energy $\epsilon^B_j$. In $H_{sr}$ we have employed the Rotating Wave Approximation (RWA) by assuming that the bandwidth of the bath is much larger than the system-reservoir coupling $\kappa_j$ such that the non-energy-conserving terms have been discarded in $H_{sr}$.

By writing down the Heisenberg equation of motions for $\ba$ and $\hat{r}_j$, and deploying the Markovian approximation for the system-reservoir coupling~\cite{Meystre2007,Scully1997}, we obtain, after a straightforward algebra, the following 
\begin{align}
\label{heisenberg}
i\frac{\partial \ba}{\partial t} = (\epsilon_\bk - i\gamma)\ba + iF_\pu e^{-i\epsilon_\pu t}\delta_{\bk,\bk_\pu} + \big[\ba, \frac{g}{2A}\sum_{\bk,\bk',\bq} \hat{b}^\dag_\bk \hat{b}^\dag_{\bk'}  \hat{b}_{\bk'-\bq} \hat{b}_{\bk+\bq}\big],
\end{align}
where the polaron-polariton decay is $\gamma = 2\pi |\kappa(\epsilon_\bk)|^2\mathcal{D}(\epsilon_\bk)$ with $\kappa(\epsilon)$ being a continuous interpolation of $\kappa_j$ and $\mathcal{D}(\epsilon)$ the density of the states of the reservoir~\cite{Scully1997}. The microscopic details of the reservoir are not important and thus for $\gamma$ we pick an experimentally realistic value of $\gamma = 0.4$ meV. Moreover, we have ignored for simplicity the quantum noise term arising from the system-reservoir coupling.

For the purpose of this work, we choose $\bk_\pu =0$. Now we assume that the polaron-polariton state at $\bk_\pu$ acquires a macroscopic population due to the pump. We therefore approximate $\hat{b}_{\bk_\pu}$ as a complex number such that $\hat{b}_{\bk_\pu} \approx \sqrt{A}\Psi_0(t)$, where $\Psi_0(t)$ is the condensation wave function. At the Gross-Pitaevskii mean-field level, the contribution of states of $\bk \neq \bk_\pu$ is neglected and from Eq. \eqref{heisenberg} we obtain for the condensed state the following:
\begin{align}
\label{gpe1}
i\partial \Psi_0(t) =  (\epsilon_0 - i\gamma) \Psi_0(t) + iF_\pu e^{-i\epsilon_\pu t} + g|\Psi_0(t)|^2\Psi_0(t),
\end{align}
where $\epsilon_0 \equiv \epsilon_{\bk=0}$. By assuming that the condensate wave function follows the time dependence of the driving field, i.e. $\Psi_0(t) = \Psi_0 e^{-i\epsilon_\pu t}$, one acquires the non-equilibrium Gross-Pitaevsii equation:
\begin{align}
\label{gpe}
\epsilon_\pu \Psi_0 = (\epsilon_0 - i\gamma) \Psi_0 + iF_\pu + g|\Psi_0|^2\Psi_0,
\end{align}
which is the same as Eq. (6) in the main text. 

The condensation wave function can be written as $\Psi_0 = \sqrt{n_0} e^{i\phi_0}$, where $n_0$ is the condensation density. In case of an equilibrium condensate, any value of $\phi_0$ would yield the same ground state energy and equally well provide a solution for the equilibrium GPE. However, it turns out that the non-equilibrium GPE of Eq.~\eqref{gpe} can be solved in general only for some specific value of $\phi_0$, i.e. the external pumping breaks explicitly the $U(1)$ phase symmetry by picking a definite complex phase $\phi_0$ for the condensation wave function.

Now we study the excitation spectrum of the non-equilibrium condensate. To this end, we write $\ba$ as
\begin{align}
\ba = \big[ \sqrt{A} \Psi_0 \delta_{\bk,0} + \delta \ba(t) \big] e^{-i\epsilon_{\pu}t},   
\end{align}
where we have introduced the fluctuation term on top of the condensate, $\delta \ba(t)$, to account for the non-condensed polaron-polaritons. To describe the fluctuations, we introduce the bosonic Green's function which in the imaginary-time domain reads
\begin{align}
\mathcal{G}_B(\bk,\tau) \equiv -\langle T_\tau   \begin{bmatrix} \delta \ba(\tau) \\ \delta \bcm(\tau) \end{bmatrix}   
  \begin{bmatrix} \delta \bc(0) & \delta \bam(0) \end{bmatrix}  \rangle = 
  \begin{bmatrix}
   -\langle T_\tau \delta\ba(\tau) \delta\bc(0) \rangle & -\langle T_\tau \delta\ba(\tau) \delta\bam(0) \rangle  \\   -\langle T_\tau \delta\bcm(\tau) \delta\bc(0) \rangle &  -\langle T_\tau \delta\bcm(\tau) \delta\bam(0) \rangle,
  \end{bmatrix}
\end{align}
where $\bk \neq 0$ as we are considering the non-condensed polaron-polaritons only. In the Matsubara frequency space one has the Dyson equation~\cite{Fetter1971}:
\begin{align}
\label{dyson}
&\mathcal{G}_B^{-1}(\bk,i\omega_n) = \mathcal{G}_{B,0}^{-1}(\bk,i\omega_n) -  \Sigma_B(\bk,i\omega_n).
\end{align}
Here $\Sigma_B$ is the self-energy arising from finite interactions between polaron-polaritons and $\mathcal{G}_{B,0}$ is the non-interacting Green's function, i.e. in the absence of interactions, $g=0$, one has $\mathcal{G}_{B} = \mathcal{G}_{B,0}$.

To proceed, we first evaluate $\mathcal{G}_{B,0}$. We denote the Hamiltonian without the interactions as $H_0 \equiv H_L(g=0)$. Then, by using the Markovian approximation and ignoring the quantum noise terms, we obtain the following imaginary-time equations of motion~\cite{Meystre2007,Scully1997}:
\begin{align}
&[\delta \ba(\tau) e^{\epsilon_\pu \tau},H_0] =  (-\epsilon_\bk + i\gamma) \delta \ba(\tau) e^{\epsilon_\pu \tau} \nonumber \\
&[\delta \bc(\tau) e^{-\epsilon_\pu \tau},H_0] =  (\epsilon_\bk + i\gamma) \delta \bc(\tau) e^{-\epsilon_\pu \tau}.
\end{align}
The latter equation would have  a wrong form for the decay part if we had used the non-hermitian form $H_{decay} = \sum_\bk i\gamma \bc \ba$ describing the decay instead of the system-reservoir formalism. Now, it is straightforward to take the time derivative of $\mathcal{G}_{B,0}(\bk,\tau)$ to show that
\begin{align}
& \partial_\tau [\mathcal{G}_{B,0}(\bk,\tau)]_{11} = -\delta(\tau) -(\epsilon_\bk -i\gamma - \epsilon_\pu)[\mathcal{G}_{B,0}(\bk,\tau)]_{11} \nonumber \\
& \partial_\tau [\mathcal{G}_{B,0}(\bk,\tau)]_{22} = \delta(\tau) + (\epsilon_\bk + i\gamma -\epsilon_\pu)[\mathcal{G}_{B,0}(\bk,\tau)]_{22}.
\end{align}
By transforming these expressions to the Matsubara frequency space, one has
\begin{align}
\label{g0}
\mathcal{G}_{B,0}^{-1}(\bk,z) = \begin{bmatrix} z - \tilde{\epsilon}_\bk + i\gamma & 0 \\ 0 & -z - \tilde{\epsilon}_\bk -i\gamma
\end{bmatrix},   
\end{align}
where $\tilde{\epsilon}_\bk \equiv \epsilon_\bk - \epsilon_\pu$.

\begin{figure}
  \centering
    \includegraphics[width=0.5\textwidth]{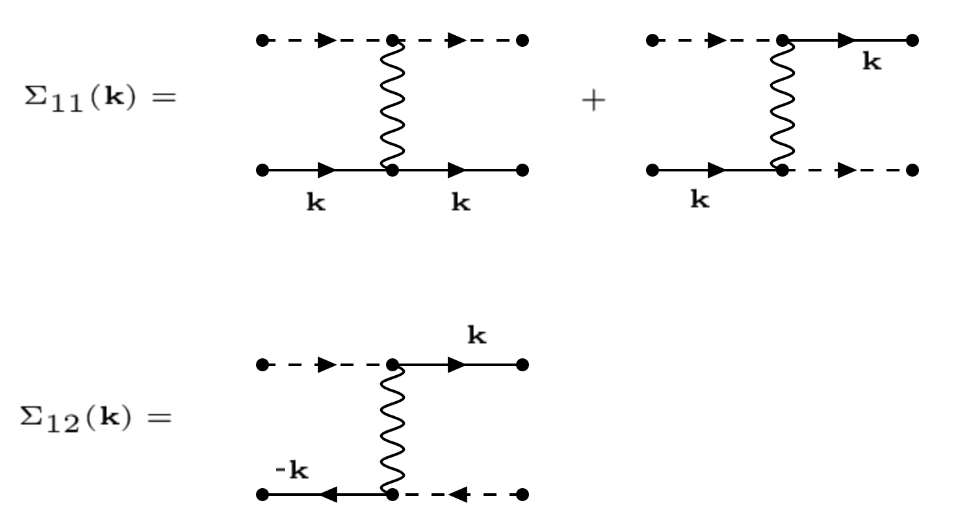}
    \caption{Self-energy diagrams included in the Bogoliubov approximation.  The solid propagators are the non-condensed polaron-polaritons, whereas the dashed lines depict polaron-polaritons propagating into and out of the condensate with the wavefunction $\Psi = \sqrt{n_0} e^{i\phi_0}$. The wavy lines depict the interaction polariton interaction $g$.}
   \label{Fig:diagrams}
\end{figure}

The self-energy $\Sigma_B$ can be in general evaluated with the diagrammatic Beliaev theory~\cite{Fetter1971}. Here, we deploy the first order Beliaev theory, i.e. the Bogoliubov theory that includes only the diagrams presented in Fig.~\ref{Fig:diagrams}. Since we are interested on low-energy states, for simplicity for the scattering between two polaritons with momenta $\mathbf k$ and $\mathbf k'$ that exchange a momentum $\bq$, we approximate the polariton-polariton interactions by $g_{\mathbf k,\mathbf k';\mathbf q}=g=g_{xx}\tilde C(\mathbf k=0)^4Z^2(\mathbf k=0)$, and based on the same argument, we assume a momentum independent damping rate $\gamma$. Then the Bogoliubov self-energy reads as
\begin{align}
\label{sigma}
\Sigma_B = \begin{bmatrix} 2gn_0  & gn_0  e^{i2\phi_0} \\ gn_0 e^{-i2\phi_0} & 2gn_0
\end{bmatrix}.   
\end{align}
By inserting the expressions \eqref{g0} and \eqref{sigma} to the Dyson equation of Eq.~\eqref{dyson}, one finally obtains the Bogoliubov Green's function used in the main text:
\begin{align}
\mathcal{G}_B^{-1}(\bk,z) = \begin{bmatrix} 
z - \tilde{\epsilon}_\bk + i\gamma -2gn_0 & -gn_0  e^{i2\phi_0} \\ -gn_0 e^{-i2\phi_0} & -z - \tilde{\epsilon}_\bk -i\gamma -2gn_0.
\end{bmatrix}    
\end{align}
Excitation spectrum is then given by $\det[\mathcal{G}_B^{-1}(\bk,E_\bk)] = 0$, yielding the dispersion relation written down in the main text. As the complex phase of condensation wave function, $\phi_0$, does not affect the excitation spectrum or the stability condition, $\Im [E_\bk] <0$, we have in the main text for simplicity set $\phi_0=0$ in the expression of $\mathcal{G}_B^{-1}(\bk,z)$.

\section{Origin of non-linear features}
In this section we point out the origin of two different non-linear regimes seen in the low and high electron density limits presented in Fig. 1(c) of the main text. We specifically show that the non-linear feature in the low-$\epsilon_F$ regime arises predominantly from the increasing blue detuning of the laser, i.e.  $\epsilon_\pu -\epsilon_0$, as a function of increasing $\epsilon_F$, whereas the high-$\epsilon_F$ non-linear feature results in from the interplay of both increasing $\epsilon_\pu -\epsilon_0$ and $g$ as a function of the electron density. 

\begin{figure}
  \centering
    \includegraphics[width=0.8\textwidth]{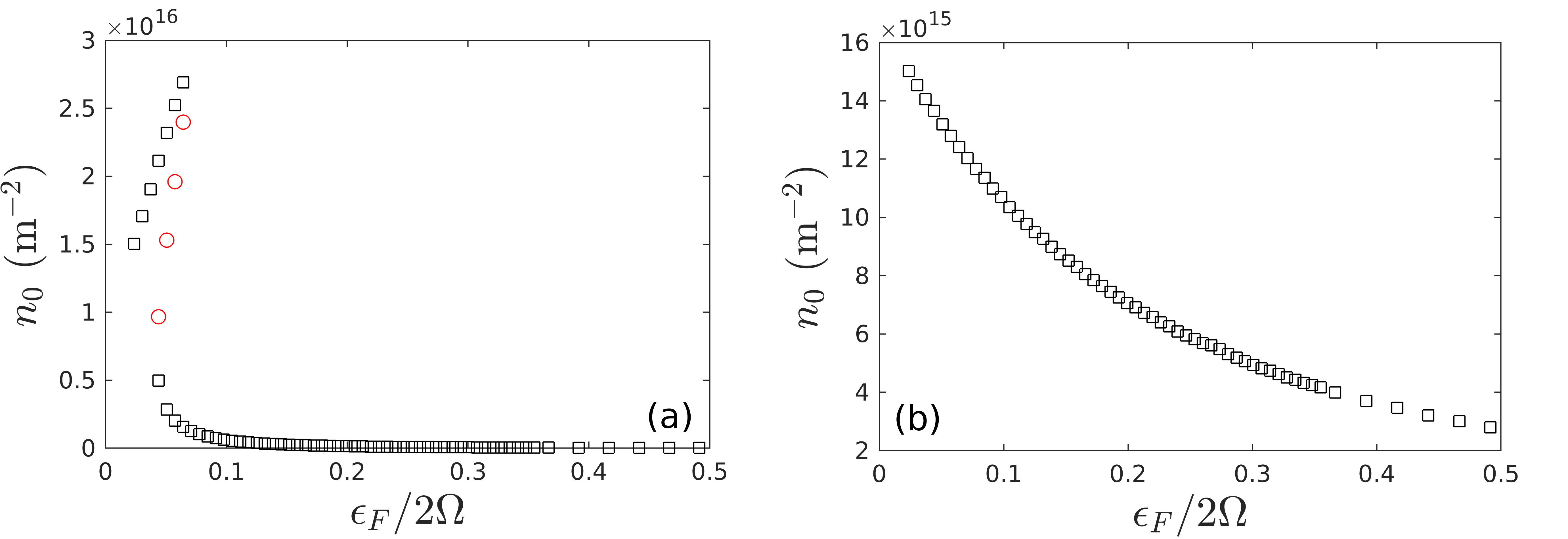}
    \caption{(a) Condensate density $n_0$ as a function of electron density for fixed $g(\epsilon_F/2\Omega = 0.023)/g_{xx} = 0.0029$. (b) Condensate density $n_0$ as a function of electron density for fixed $\epsilon_0(\epsilon_F/2\Omega = 0.023)/2\Omega = -2.224$. Black square (red dots) indicate stable (unstable) solutions for the non-equilibrium GPE. The pump intensity and energy are the same as in Figs.1(b)-(c) of the main text.}
   \label{fig:supp3}
\end{figure}

To demonstrate these claims, we performed two computations: one with fixed $\epsilon_0$ and one with fixed $g$ such that in both the computations either only $\epsilon_0$ or $g$ was allowed to change as a function of $\epsilon_F$. The results are shown in Fig.~\ref{fig:supp3}: in Fig.~\ref{fig:supp3}(a) we show $n_0$ as a function of $\epsilon_F$ by fixing the interaction $g$, whereas in Fig.~\ref{fig:supp3}(b) the polaron-polariton ground state energy $\epsilon_0$ is held fixed. The pump intensity and frequency are the same as in Fig. 1(c) of the main text. In Fig.~\ref{fig:supp3}(a) [Fig.~\ref{fig:supp3}(b)] the value of interaction $g$ (ground state energy $\epsilon_0)$ is given by the lowest value of $\epsilon_F$, i.e. $g/g_{xx} = 0.0029$ ($\epsilon_0/2\Omega= 0.023$).

From Fig.~\ref{fig:supp3}(a) one observes that despite having a constant value for $g$, a non-linear behavior of three solutions can still be produced at the low-$\epsilon_F$ regime, in a similar fashion as in Fig. 1(c) of the main text. In contrast, with fixed $\epsilon_0$ this feature is absent, see Fig.~\ref{fig:supp3}(b). As a consequence,
the non-linear behavior seen at the low-$\epsilon_F$ regime in Fig. 1(c) of the main text arises due to the modification of $\epsilon_0$ as a function of $\epsilon_F$.

From Fig.~\ref{fig:supp3}(a) we also see that for fixed $g$ one cannot obtain a non-linear behavior at larger $\epsilon_F$, in contrast to the feature seen in Fig. 1(c) of the main text. Furthermore, the non-linearity is also absent with fixed $\epsilon_0$, see Fig.~\ref{fig:supp3}(b). We can thus conclude that the non-linear behavior at the high-$\epsilon_F$ regime in Fig. 1(c) of the main text arises because both $\epsilon_0$ and $g$ depend strongly on the electron density. Thus, such a non-linear feature cannot be achieved by simply tuning the pump energy in the absence of the 2DEG.

%